\documentclass[floats,aps,amsmath,showpacs, twocolumn]{revtex4}
\usepackage{amsmath}
\usepackage{graphicx}
\usepackage{amsfonts}

\newcommand{\be}{\begin{equation}}
\newcommand{\ee}{\end{equation}}
\newcommand{\ba}{\begin{eqnarray}}
\newcommand{\ea}{\end{eqnarray}}

\newcommand{\opone}{\leavevmode\hbox{\small1\kern-3.8pt\normalsize1}}

\begin{document}

\title{Periodic-Orbit Theory of Level Correlations}
\author{Stefan Heusler$^1$, Sebastian M{\"u}ller$^2$, Alexander Altland$^3$,
Petr Braun$^{1,4}$, Fritz Haake$^1$}

\address{$^1$Fachbereich Physik, Universit{\"a}t Duisburg-Essen,
  47048 Duisburg, Germany\\$^2$ Cavendish Laboratory, University of
  Cambridge, Cambridge CB30HE, UK\\
  $^3$Institut f{\"u}r Theoretische Physik, Z{\"u}lpicher Str 77, 50937 K{\"o}ln,
  Germany\\$^4$Institute of Physics, Saint-Petersburg University,
  198504 Saint-Petersburg, Russia}

\begin{abstract}
  We present a semiclassical explanation of the so-called
  Bohigas-Giannoni-Schmit conjecture which asserts universality of
  spectral fluctuations in chaotic dynamics. We work with a generating
  function whose semiclassical limit is determined by quadruplets of
  sets of periodic orbits. The asymptotic expansions of both the
  non-oscillatory and the oscillatory part of the universal spectral
  correlator are obtained. Borel summation of the series reproduces
  the exact correlator of random-matrix theory.
\end{abstract}

\pacs{05.45.Mt, 03.65.Sq}
\maketitle

Quantum spectra of \textit{individual} chaotic systems can be
phenomenologically described in terms of random-matrix
theory (RMT) \cite{Stoeckmann,Haake}. This universality -- asserted by
the celebrated \textit{\ Bohigas-Giannoni-Schmit conjecture} (BGS)
\cite{BGS} -- is an empirical fact, supported by a huge body of
experimental and numerical data. Proving its conceptual origin remains
one of the fundamental challenges in quantum or wave chaos.

Spectral fluctuations are conveniently characterized in terms of the
two-point correlation function, $R(\epsilon )=\Delta ^{2}\left\langle \rho
(E+\frac{\epsilon \Delta }{2\pi })\rho (E-\frac{\epsilon \Delta }{2\pi }%
)\right\rangle -1$, where $\rho (E)$ is the energy-dependent density of
states, $\Delta \equiv 1/\langle \rho \rangle $ the mean level spacing and $\langle \cdot \rangle$
denotes averaging over the energy $E$. Predictions made by RMT for the
two-point correlation function are fully universal in that they depend
only on the parameter $\epsilon $, and the fundamental symmetries of the
system under consideration. Specifically,  the
complex representation $R(\epsilon )=\lim_{\gamma \to 0}\text{Re}
\;C(\epsilon ^{+})$ where $\epsilon^\pm=\epsilon\pm i\gamma$ and $C(\epsilon ^{+})=\frac{\Delta ^{2}}{2\pi ^{2}}%
\langle \mathrm{tr}\,G^{+}(E+\epsilon \Delta /2\pi )\,\mathrm{tr} \,G^{-}(E-\epsilon \Delta /2\pi
)\rangle -\frac{1}{2}$ is employed, with $G^{\pm }(E)=(E\pm i\gamma\Delta/2\pi-H)^{-1}$
and $H$ the Hamiltonian. The Wigner--Dyson unitary and orthogonal
symmetry classes (the only ones to be considered here) of RMT afford
the asymptotic series
\begin{eqnarray}\label{C}
C(\epsilon ^{+})\sim
\begin{cases}
\frac{1}{2(i\epsilon ^{+})^{2}}-\frac{\mathrm{e}^{2i\epsilon ^{+}}}{%
2(i\epsilon ^{+})^{2}} & \mbox{unitary} \\
\frac{1}{(i\epsilon ^{+})^{2}}+\sum_{n=3}^{\infty }\frac{(n-3)!(n-1)}{%
2(i\epsilon ^{+})^{n}} &  \\
+\,\mathrm{e}^{2i\epsilon ^{+}}\sum_{n=4}^{\infty }\frac{(n-3)!(n-3)}{%
2(i\epsilon ^{+})^{n}} & \text{orthogonal}\;.%
\end{cases}
\end{eqnarray}
In either case, $C(\epsilon ^{+})$ is a sum of a non-oscillatory part
(power series in $1/\epsilon ^{+}$) and an oscillatory one
($\mathrm{e}^{2i\epsilon ^{+}}$ times a series in $1/\epsilon ^{+}$).
Resumming the series by Borel summation techniques and extrapolating
to small positive values of $\epsilon $, one obtains
$R(\epsilon)+1\propto\epsilon^{\beta }$ for $\epsilon\to0$, a
signature of the level repulsion symptomatic for chaos ($\beta =1,2$
for the case of orthogonal, unitary symmetry, resp.)

The question to be addressed below is how to obtain the RMT prediction
(\ref{C}) for a concrete chaotic (globally hyperbolic) quantum system.
A step in this direction was recently made~\cite{EssenFF} on the basis
of Gutzwiller's semiclassical periodic-orbit theory \cite{Gutzwiller}.
Gutzwiller represents the level density $\rho (E)$ as a sum over
periodic orbits, whereupon the function $R(\epsilon )$ becomes a sum
over orbit pairs \cite%
{Berry,Argaman,SR}. Relevant contributions to that double sum were
shown \cite{EssenFF} to originate from orbit pairs which are
identical, mutually time-reversed, or differ only by connections in
certain close self-encounters. By summing over all distinct families
of orbit pairs, the Fourier transform of $R(\epsilon )$, the spectral
form factor $K(\tau )$, was found to coincide with the RMT prediction
for times $t=\tau T_{H}$ smaller than the Heisenberg time $T_{H}=2\pi
\hbar /\Delta $, the time needed to resolve the mean level spacing.
The behavior of $K(\tau )$ for $\tau >1$, also known from RMT, was
left unexplained.

We now want to fill the gap left. As results we will
obtain the full expression for the correlation function in the case of unitary
symmetry, and an asymptotic $1/\epsilon $-expansion amenable to Borel
summation in the orthogonal case. The oscillatory term, Fourier
transformed, then complements $K(\tau)$ to its full
form at $\tau>1$. In many respects, our reasoning is inspired by the
field theoretical formulation of RMT correlation functions~\cite{AIE},
notably the existence of \textquotedblleft anomalous saddle
points\textquotedblright\ in the nonlinear $\sigma $-model \cite{Kamenev}.
It also affords a new interpretation of ideas underlying the
``bootstrapping'' \cite{Bootstrap}.

The basic idea of our approach is to consider representations of
$C(\epsilon^+) $ different from the standard one in terms of the product
of a single retarded and advanced Green function. We start from the
generating function
\begin{equation}  \label{Z}
Z=\left\langle \frac{\det(E_C^+-H)\det(E_D^--H)}{\det(E_A^+-H)\det(E_B^--H)}
\right\rangle\,
\end{equation}
where $E_{A,B,C,D}^\pm$ are energies in the vicinity of $E$ defined by $%
E_{A,B,C,D}^\pm=E+\epsilon_{A,B,C,D}^\pm \Delta/2\pi$. From $Z$, the complex
correlator can be accessed as
\begin{equation}  \label{deriv}
\lim_{\gamma\to0} C(\epsilon^+)=-\frac{1}{2}+2\frac{\partial^2 Z}{\partial
\epsilon_{A}^+\partial \epsilon_{B}^-}\big|_{\|,\times}\;.
\end{equation}
The two derivatives produce the product $({\rm tr}G^+)({\rm tr}G^-)$
under the energy average. If we subsequently identify the energies
``columnwise'' ($\|$): $ \epsilon_A^+=\epsilon_C^+={\epsilon^+}$,
$\epsilon_B^-=\epsilon_D^-=-{ \epsilon^+}$, or
``crosswise''($\times$): $\epsilon_A^+={\epsilon^+}$, $
\epsilon_B^-=-{\epsilon^+}$, $\epsilon_C^+=-{\epsilon^-}$,
$\epsilon_D^-={ \epsilon^-}$, $\gamma\to0$ the ratio of determinants
approaches unity. The first representation for $C(\epsilon^+)$ does
not even require the limit $\gamma\to0$; it is widely used in RMT. The
second representation is crucial for us. Importantly, the
\textit{semiclassical approximation} of either of these two exact
representations misses contributions to $Z$, and therefore to
$C(\epsilon^+)$: the first representation yields only the
non-oscillatory contributions, and the second (without
''$-\frac{1}{2}$'') only the oscillatory ones; adding both we will
recover the universal two-point correlator.

To see the emergence of these structures, let us represent the
determinants in (\ref{Z}) as
\begin{eqnarray}
&&\det \,(E_{A}^{+}-H)^{-1}=\exp\Big\{ -\int^{E_{A}^+}dE\,\mathrm{tr}
\,G^{+}(E)\Big\}  \notag  \label{orbits} \\
&\sim &\mathrm{const}\times \exp \Big(i\pi E_{A}^{+}/\Delta +\sum_{a}F_{a}
\mathrm{e}^{\frac{i}{\hbar }S_{a}(E_{A}^{+})}\Big)\,,
\end{eqnarray}%
where the last line invokes the semiclassical expansion of the
integrated Green function into a smooth (Weyl) average and a
fluctuating (Gutzwiller) part; the latter is a sum of periodic orbits
$a$ with action $S_{a}$ and stability amplitude $F_{a}$; for
simplicity, we assume the average level density $1/\Delta$ to be
constant; the periodic-orbit sum converges for $\mathrm{Im}E_{A}^{+}$
large enough; the \textquotedblleft const\textquotedblright\ in
(\ref{orbits}) comes from the lower limit of the energy integral and
cancels from the ratio of determinants in $Z$.

Expanding the exponential in (\ref{orbits}) we get a sum over
non-ordered sets of orbits $A$ (\textquotedblleft
pseudo-orbits\textquotedblright),
\begin{equation}\label{det}
\det (E_{A}^{+}-H)^{-1}\!\!\sim \!\mathrm{const}\times \mathrm{e}^{i\pi
E_{A}^{+}/\Delta}\sum_{A}\!F_{A}\mathrm{e}^{\frac{i}{\hbar }
S_{A}(E_{A}^{+})}\,.
\end{equation}
A pseudo-orbit $A$ may involve any number $n_{A}$ of component orbits
($n_{A}=0$ pertains to the empty set which contributes unity to the
sum); $F_{A}$ is the product of the stability amplitudes and $S_{A}$
the cumulative action of all component orbits. Expressing all four
determinants in (\ref{Z}) similarly to (\ref{orbits},\ref{det})
(e.~g., using $\det (E_{B}^{-}-H)=(\det (E_{B}^{+}-H))^{\ast }$) and
writing $S(E+\epsilon \Delta /2\pi )\sim S(E)+T(E)\epsilon \Delta
/2\pi \;$ ($T$ is the period of an orbit, or the sum of periods in a
pseudo-orbit) we approximate the generating function as
\begin{eqnarray}
Z\;\sim \!\!\!\! &&\mathrm{e}^{\frac{i}{2}(\epsilon _{A}^{+}-\epsilon
_{B}^{-}-\epsilon _{C}^{+}+\epsilon _{D}^{-})}  \label{Zsc1} \\
&&\hspace{-0.7cm}\times \big\langle\textstyle{\exp \big(\sum_{a}F_{a}\mathrm{
e}^{\frac{i}{\hbar }S_{a}(E)+i\frac{T_{a}}{T_{H}}\epsilon
_{A}^{+}}+\sum_{b}F_{b}^{\ast }\mathrm{e}^{-\frac{i}{\hbar }S_{b}(E)-i\frac{
T_{b}}{T_{H}}\epsilon _{B}^{-}}}  \notag \\
&&\!\!-\textstyle{\sum_{c}F_{c}\,\mathrm{e}^{\frac{i}{\hbar }S_{c}(E)+i\frac{
T_{c}}{T_{H}}\epsilon _{C}^{+}}-\sum_{d}F_{d}^{\ast }\mathrm{e}^{-\frac{i}{
\hbar }S_{d}(E)-i\frac{T_{d}}{T_{H}}\epsilon _{D}^{-}}\big)\big\rangle}
\notag \\
\!\! &=&\mathrm{e}^{\frac{i}{2}(\epsilon _{A}^{+}-\epsilon _{B}^{-}-\epsilon
_{C}^{+}+\epsilon _{D}^{-})}\big\langle\!\!\sum_{A,B,C,D}\!\!F_{A}F_{B}^{
\ast }F_{C}F_{D}^{\ast }(-1)^{n_{C}+n_{D}}  \notag \\
&&\hspace{2cm}\times \;\mathrm{e}^{i(S_{A}(E)-S_{B}(E)+S_{C}(E)-S_{D}(E))/
\hbar }  \label{Zsc2} \\
&&\hspace{2cm}\times \;\mathrm{e}^{i(T_{A}\epsilon _{A}^{+}-T_{B}\epsilon
_{B}^{-}+T_{C}\epsilon _{C}^{+}-T_{D}\epsilon _{D}^{-})/T_{H}}\big\rangle\;.
\notag
\end{eqnarray}
Here, the mean density produces a phase factor $\mathrm{\
  e}^{\frac{i}{2} (\epsilon _{A}^{+}-\epsilon _{B}^{-}-\epsilon
  _{C}^{+}+\epsilon _{D}^{-})}$.  When representing the correlator as
in (\ref{deriv}), that phase factor turns into 1 and
$\mathrm{e}^{2i\epsilon }$ for the columnwise and crosswise
identifications of energies, respectively. Indeed, then, we can
recover either the non-oscillatory or the oscillatory contributions to
$C(\epsilon ^{+})$.

Another phase factor involves the difference $\Delta S\equiv
S_A(E)-S_B(E)+S_C(E)-S_D(E)$ between the cumulative actions of $(A,C)$
and $ (B,D)$. Due to this factor, systematic contributions in the
limit $\hbar\to 0 $ can arise only for quadruplets of pseudo-orbits whose
action difference is of the order of $\hbar$ or smaller.

The most basic of quadruplets have each of the component orbits of $A$
and $C$ repeated in either $B$ or $D$, such that $ \Delta S=0$. These
\textquotedblleft diagonal quadruplets\textquotedblright\ may be
summed by a lowest-order cumulant expansion: denoting the
periodic-orbit sums in the exponent of (\ref{Zsc1}) by $X$, we may
write $\left\langle \mathrm{e} ^{X}\right\rangle_{\rm diag} = \exp
\{\left\langle X^{2}\right\rangle_{\rm diag} /2\}$, wherein
$\left\langle X^{2}\right\rangle_{\rm diag}$ contains only pairs of
identical orbits. We obtain
\begin{eqnarray}
Z_{\mathrm{diag}}\sim \hspace{-0.3cm} &&\exp \textstyle{{\left\langle
X^{2}\right\rangle _{\mathrm{diag}}/2}}  \label{Zdiag0} \\
\, &\sim &\hspace{-0.2cm}\textstyle{\exp \big\langle\sum_{a}|F_{a}|^{2}\big(
\mathrm{e}^{i\frac{T_{a}}{T_{H}}(\epsilon _{A}^{+}-\epsilon _{B}^{-})}-
\mathrm{e}^{i\frac{T_{a}}{T_{H}}(\epsilon _{A}^{+}-\epsilon _{D}^{-})}}\big)
\notag \\
&&\hspace{0.6cm}\textstyle{-\sum_{c}|F_{c}|^{2}\big(\mathrm{e}^{i\frac{T_{c}
}{T_{H}}(\epsilon _{C}^{+}-\epsilon _{B}^{-})}-\mathrm{e}^{i\frac{T_{c}}{
T_{H}}(\epsilon _{C}^{+}-\epsilon _{D}^{-})}\big)\big\rangle}.  \notag
\end{eqnarray}
Relying on ergodicity, the resulting sums over orbits may be evaluated
by the sum rule of Hannay and Ozoria de Almeida \cite{HOdA},
$\sum_{a}|F_{a}|^{2}(\cdot )\approx \int_{T_{0}}^{\infty
}\frac{dT}{T}(\cdot )$; the lower limit of the integration is a
minimal period $T_{0}$ starting from which periodic orbits behave
approximately ergodically. By this rule, e.g., the first sum turns
into $-\ln (i(\epsilon _{A}^{+}-\epsilon
_{B}^{-}))+\mathrm{const}+\mathcal{\ O}(\hbar )$. All four sums yield
\begin{equation}\label{diag1}
Z_{\mathrm{diag}}\sim \mathrm{e}^{\frac{i}{2}(\epsilon _{A}^{+}-\epsilon
_{B}^{-}-\epsilon _{C}^{+}+\epsilon _{D}^{-})}\left( \frac{(\epsilon
_{A}^{+}-\epsilon _{D}^{-})(\epsilon _{C}^{+}-\epsilon _{B}^{-})}{(\epsilon
_{A}^{+}-\epsilon _{B}^{-})(\epsilon _{C}^{+}-\epsilon _{D}^{-})}\right)
^{\kappa },
\end{equation}
with $\kappa =1$ for the unitary class. For the orthogonal class we must
also consider pairs of mutually time reversed orbits. Therefore, each
sum in (\ref{Zdiag0}) must be multiplied by 2 whereupon in the final
result (\ref {diag1}) we have $\kappa =2$.

Taking derivatives and columnwise identified energies, we recover the
leading non-oscillatory contribution to the two-point correlator,
$(i\epsilon^+)^{-2}/\beta$. Crosswise identified energies yield the oscillatory
contribution $-\mathrm{\ e}^{2i\epsilon}(i\epsilon)^{-2}/2$ for $\beta=2$ (thus
completely reproducing (\ref{C})), while for $\beta=1$ we get zero,
i.~e., no oscillatory term arises up to $\mathcal{O}(\epsilon^{-2})$,
again as in RMT.

\begin{figure}[tbp]
\begin{center}
\includegraphics[scale=0.6]{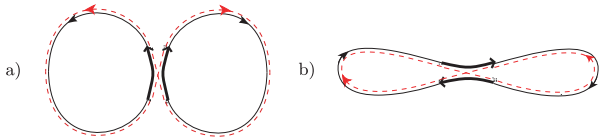}
\end{center}
\caption{``Diagrams'' of simplest
  (pseudo-)orbits in configuration space. Arrows indicate sense of
  traversal.}
\label{bild}
\end{figure}

Going beyond the above level of approximation, we note that small
phases may also arise from component orbits $B$ and $D$ differing from
$A$ and $C$ in \textit{topology}, but only weakly in action. By way of
example, consider a pseudo-orbit $A$ with just one component, while
$C$ is the empty set. Assume that $A$ contains a so-called
\textquotedblleft 2-encounter\textquotedblright , where two stretches
of the orbit come close in phase space. This situation is depicted in
Fig.~\ref{bild}a (where the phase-space encounter contains a crossing
in configuration space). The two stretches start and end at altogether
four \textquotedblleft ports\textquotedblright , and are connected to
each other through two external \textquotedblleft
links\textquotedblright\cite{foot} .  Changing the way the ports are
connected inside the encounter, as illustrated by the dashed line in
Fig.~\ref{bild}a, we obtain two disjoint periodic orbits, each
containing one link and one encounter stretch. The existence proof of
the latter two orbits as exponentially close to the links of the
original orbit requires chaos and was in essence given in
Ref.~\cite{SR}. These two orbits can either both be included in $B$
such that $D$ remains empty, or vice versa, or one is included in $B$
and the other one in $D$. At any rate the cumulative actions of
$(A,C)$ and $(B,D)$ almost coincide.

A second example, a so-called Sieber/Richter pair \cite{SR}, is shown
in Fig.~\ref{bild}b: Here two orbits differ in an encounter wherein
two stretches are approximately mutually time-reversed. One of these
orbits is included in $A$ leaving $C$ empty or vice versa, whereas the
other is included in $B$ or $D$. Such orbit pairs exist only in
time-reversal invariant systems since we assume that the sense of
motion on an encounter stretch or a link can be reverted in time.

More complicated quadruplets involve any number of orbits, and $(A,C)$
and $ (B,D)$ can differ in any number of encounters where arbitrarily
many stretches come close in phase space (modulo time reversal for
time-reversal invariant systems); encounters may be self-encounters
within periodic-orbit components in a pseudo-orbit, mutual encounters
of different periodic orbits within one pseudo-orbit, or even from
different pseudo-orbits.

As in the diagonal approximation some of the component orbits of
$(A,C)$ may just be repeated in $(B,D)$. To evaluate the generating
function, we must sum over all quadruplets of this type. We can split
the sum into one over the ``diagonal'' parts of these quadruplets and
one over the orbits differing in encounters. The first subsum yields
$Z_{\mathrm{diag}}$ as in (\ref{diag1}) such that
$Z=Z_{\mathrm{diag}}(1+Z_{\mathrm{off}})$ with
\begin{eqnarray}  \label{off}
Z_{\mathrm{off}}=\sum_{A,B,C,D\atop{\text{diff. in enc.}}}&&\hspace{-.4cm}
\Big\langle F_{A} F_{B}^* F_{C} F_{D}^*(-1)^{n_{C}+n_{D}} \\
&&\hspace{-.4cm}\times\, \mathrm{e}^{i\Delta S/\hbar} \mathrm{e}
^{i(T_{A}\epsilon_A^+-T_{B}\epsilon_B^-+T_{C}\epsilon_C^+-T_{D}
\epsilon_D^-)/T_H}\Big\rangle\,.  \notag
\end{eqnarray}

To find $Z_{\mathrm{off}}$ we must classify and count quadruplets with
encounters. Their topological ``structure'' must be dealt with
first. To that end we number all encounter stretches (stretches, for
short) of $A$ and $C$. Starting with an arbitrary stretch in the orbit
with the smallest number of stretches, we number the stretches in that
orbit in their order of traversal; we continue with the orbit with the
second-smallest number of stretches, etc. Each structure now
corresponds to one way of (i) grouping these numbered stretches into
encounters, (ii) choosing their mutual orientation if the system is
time-reversal invariant, (iii) changing connections inside the
encounters, and (iv) dividing the original orbits among $A$ and $C$
and the reconnected orbits among $B$ and $D$. Quadruplets of
pseudo-orbits can typically be described by several equivalent
structures: If we denote by $\omega_\mu$ the number of component
orbits of $%
A,C$ with $\mu$ encounter stretches, there are $\prod_\mu
\mu^{\omega_\mu}\omega_\mu!$ ways of choosing first stretches, and
ordering the $\omega_\mu$ orbits with fixed $\mu$. When summing over
structures, we must divide out that overcounting factor.

Next, pseudo-orbit quadruplets are characterized by phase-space
separations between the encounter stretches. To measure separations
for an $l$-encounter, we introduce a Poincar{\'{e}} surface of section
orthogonal to the original orbit in an arbitrary reference point in
one of the stretches.  The other stretches pierce through the same
section in $(l-1)$ further points; their phase-space separation from
the first piercing can be decomposed into components $u_{i}$ and
$s_{i}$ along the unstable and stable manifolds.  As shown in
Ref.\cite{EssenFF}, the encounter contributes to the action difference
as $\Delta S=\sum_{j}s_{j}u_{j}$ and has a duration
$t_{\mathrm{enc}}=\frac{1}{\lambda }\ln
\frac{c^{2}}{\max_{j}|s_{j}|\times\max_{k}|u_{k}|}$ where $\lambda $ is the
Lyapunov exponent and $c$ a constant whose value is unimportant.

The sum over $A,B,C,D$ in (\ref{off}) can be written as a sum over
structures and an integral over $s,u$ and the link durations $t$. The
measure to be used \cite{EssenFF,Transport} obtains a factor $1/\Omega
^{l-1}t_{\mathrm{enc}}$ from each $l$-encounter, with $\Omega $ the volume
of the energy shell. The factor $1/\Omega ^{l-1}$ arises from ergodicity
and gives the uniform probability density for the $l-1$ later
piercings to have given stable and unstable differences from the
reference one; the factor $1/t_{\mathrm{enc}}$ compensates an
overcounting due to the fact that the Poincar{\'{e}} section may be
placed anywhere inside the encounter.

We next split the phase-space integral into factors representing the
links and the encounters. To do so, we write the time $T_{A}$ as a sum
of durations of all links and encounter stretches which belong to $A$
before reconnection; $T_{B}$, $T_{C}$, and $T_{D}$ are decomposed
similarly.  We then obtain an integral $\int_{0}^{\infty
}dt\;\mathrm{e}^{it(\epsilon _{\text{$A$ or $C$} }^{+}-\epsilon
  _{\text{$B$ or $D$}}^{-})/T_{H}}$ for each link (belonging to $A$ or
$C$ before reconnection and to $B$ or $D$ afterwards) and an integral
$\int\!d^{l-1}\!sd^{l-1}\!u\frac{1}{\Omega
  ^{l-1}t_{\mathrm{enc}}}\!\mathrm{\ e}^{ \frac{i}{\hbar
  }\sum_{j}\!s_{j}u_{j}}\mathrm{e}^{i(l_{A}\epsilon
  _{A}^{+}-l_{B}\epsilon _{B}^{-}+l_{C}\epsilon _{C}^{+}-l_{D}\epsilon
  _{D}^{-})\frac{t_{\mathrm{enc}}}{T_{H}}}$ for each encounter (with $
l_{A},l_{B},l_{C},l_{D}$ the numbers of stretches of the encounter
belonging to $A,B,C,D$, and $l_{A}+l_{C}=l_{B}+l_{D}$.) Evaluating
these integrals as in \cite{Transport} we obtain a factor $i(\epsilon
_{\text{$A$ or $C$} }^{+}-\epsilon _{\text{$B$ or $D$}}^{-})^{-1}$ for
each link a factor $ i(l_{A}\epsilon _{A}^{+}-l_{B}\epsilon
_{B}^{-}+l_{C}\epsilon _{C}^{+}-l_{D}\epsilon _{D}^{-})$ for each
encounter, while $T_{H}$ cancels out. In this way, $Z_{\mathrm{off}}$
becomes the sum over structures
\begin{equation}\label{zoff}
Z_{\mathrm{off}}\!\sim \!\!\!\sum_{\mathrm{struct}}\textstyle{\!\!\frac{
\kappa ^{n_{B}+n_{D}}\prod_{\mathrm{enc}}i(l_{A}\epsilon
_{A}^{+}-l_{B}\epsilon _{B}^{-}+l_{C}\epsilon _{C}^{+}+l_{D}\epsilon
_{D}^{-})}{\!(-1)^{n_{C}+n_{D}}\!\prod_{\mu }\mu ^{\omega _{\mu }}\omega
_{\mu }!\!\prod_{\mathrm{links}}(-i)(\epsilon _{\text{$A$ \!or \!$C$}
}^{+}\,-\,\epsilon _{\text{$B$ \!or \!$D$}}^{-})}},
\end{equation}
where for $\beta =1$ the factor $\kappa ^{n_{B}+n_{D}}$ accounts for the two
different senses of motion on the \textquotedblleft
reconnected\textquotedblright\ orbits. Summarily referring to the
linear combinations of the $\epsilon _{A,B,C,D}^{\pm }$ in the link and
encounter factors as $\epsilon $, we infer that $Z_{\mathrm{off}}$ is a
power series in $1/\epsilon $. The term $(1/\epsilon )^{m}$ is provided by all
structures with $m=L-V$, with $V$ the number of encounters and $L$ the
number of stretches in a structure; note $L>V$. This remark allows to
draw all \textquotedblleft diagrams\textquotedblright\ contributing to
each of the first few orders of the expansion and to evaluate their
contributions.

For instance, the order $m=1$ is determined by the two diagrams in
Fig.~\ref{bild}, whereas for $m=2$ we need quadruplets with two
2-encounters or one 3-encounter. In the unitary case, all these
quadruplets either yield vanishing contributions to $Z$ (after summing
over all possible assignments of orbits to $A,B,C,D$) or mutually
cancel. Reassuringly, the low-order analysis of the general expression
above complies with the fact that for $\beta =2$ the diagonal
approximation exhausts the RMT result.

In the orthogonal case, off-diagonal contributions remain and the
non-oscillatory and the oscillatory parts of the correlator
$C(\epsilon )$ are obtained according to (\ref{deriv}). Not
surprisingly, the non-oscillatory terms are determined only by
\textit{\ pairs of orbits} (such as Fig.~\ref{bild}b) known from the
previous work on the small-time form factor; in the present language,
either $A$ or $C$, and either $B$ or $D $ are empty. All genuine
pseudo-orbits end up contributing nothing. The first oscillatory term,
$\propto \mathrm{e}^{2i\epsilon }/\epsilon ^{4}$, does involve
non-trivial pseudo-orbit quadruplets. It can be attributed to
quadruplets of orbits involving two Sieber/Richter pairs; further
(mutually canceling) diagrams are archived in an appendix. Proceeding to all orders we get the full asymptotic expansions (\ref{C}) \`a la RMT.

The form factor is finally obtained from the complex correlator as
$K(\tau)= \frac{1}{2\pi}\int_{-\infty+i a}^{\infty+ia}\mathrm{\
  e}^{-i2\epsilon\tau}C( \epsilon)d\epsilon, \quad a>0$. Substitution
of our expansion (\ref{C}) for $ C(\epsilon)$ and term-by-term
integration produces, for both symmetry classes, two analytic
functions different from zero for $\mathrm{Re}\, \tau>0 $ and
$\mathrm{Re}\, \tau>1$, respectively, which sum up to the RMT form
factor. The Laplace transform of the result reproduces the exact
complex correlator of RMT as $C(\epsilon)=2\int_{0}^\infty \mathrm{\
  e} ^{i2\tau\epsilon}K(\tau)d\tau$ including the power law of its
real part for $ \epsilon\to 0$, i.e., level repulsion. For conditions
under which such a back-and-forth Laplace transformation (Borel
summation) uniquely restores a function from its asymptotic series see
\cite{Sokal}.

We conclude with the following remarks. Implicit to our present
analysis is a specific order of two limits. These are the
semiclassical limit which brings in the periodic-orbit sum {\`a} la
Gutzwiller in (\ref{orbits}), alluded to as $\lim_{\Delta\to 0}$
below, and the vanishing of the imaginary part ${\rm
  Im}E^+=\gamma\Delta/2\pi$ of the complex energies and of ${\rm
  Im}\epsilon^+=\gamma$, i.~e.~$\lim_{\gamma \to 0}$. We need the
condition $\gamma\Delta/\Delta=\gamma>1$ to make the periodic-orbit
contributions to our asymptotic expansions well defined. It is worth
noting that this condition effectively limits the orbit periods as
$T<T_H$; see Eq.~(\ref{off}) where the final exponential includes a
damping ${\rm e}^{-\gamma(T_A+T_B+T_C+T_D)/T_H}$. In our limit
sequence (first $\Delta\to 0$ with $\gamma>1$ fixed, then $\gamma\to 0$) the
two representations, $\times$ and $\parallel$, become {\it
  in}equivalent, and resolve different parts of the two-point function
(oscillatory and non-oscillatory); separate asymptotic expansions for
\textit{both} are required to recover the full information. This
complementarity phenomenon is reflected in the structure of the field
theoretical approach to spectral statistics as well: the functional
integral representation of $Z$ in RMT is controlled by two saddle
points~\cite{Kamenev}. In the limit $\gamma\ll1$, both saddles equally
contribute and give $C$ in full in either representation, $\parallel$
and $\times$; one saddle provides the non--oscillatory part of $C$,
the other the oscillatory part.  However, for $\gamma > 1$, the
oscillatory part gets exponentially suppressed as ${\rm e}^{-\gamma}$
in the $\parallel$-representation, while the $\times$-representation
has the non-oscillatory part exponentially damped. The perturbative
$\epsilon^{-1}$--expansions of the surviving parts, on the other hand,
quantitatively agree with the results of our semiclassical analysis,
to all orders in $1/\epsilon$.  We shall further discuss analogies
semiclassics/field theory in a forthcoming paper.

We thank the SFB/TR12 of the Deutsche Forschungsgemeinschaft and
the EPSRC for funding and Ben Simons and Hans-J{\"u}rgen Sommers
for fruitful discussions.

\newpage

\appendix{\it Appendix: Leading contributions}

To illustrate our results, we evaluate the contributions of the
simplest quadruplets of pseudo-orbits. We start with the "diagram"
in Fig. 1a, where an orbit involving an encounter of two almost
parallel stretches decomposes into two disjoint partner orbits.
First, assume that the initial orbit is included in the
pseudo-orbit $A$. Each of the two partner orbits can be included
either in $B$ or in $D$. This leads to four different
possibilities, whose contributions to $Z_{\rm off}$ can be evaluated using the
diagrammatic rules introduced above,

\begin{eqnarray}
\frac{2 i \epsilon_A^+ - 2 i \epsilon_B^-}{( - i (\epsilon_A^+ - \epsilon_B^-))^2}
- 2 \frac{2 i \epsilon_A^+ -  i \epsilon_B^- - i \epsilon_D^-}{( - i (\epsilon_A^+
- \epsilon_B^-))(- i (\epsilon_A^+ - \epsilon_D^-))}
\\ \nonumber
+ \frac{2 i \epsilon_A^+ - 2 i \epsilon_D^-}{( - i (\epsilon_A^+ -
\epsilon_D^-))^2}   = 0\,.
\end{eqnarray}

In a similar way, all possibilities with the initial orbits
included in $B, C$ or $D$ also sum to zero. The  orbit quadruplets
corresponding to the "diagram" in Fig. 1a thus yields no net
contribution.

Next, we consider Sieber/Richter pairs as in Fig. 1b. These pairs
require time-reversal invariance and therefore exist only in the
orthogonal case. According to our diagrammatic rules,
Sieber/Richter pairs where, say, the original orbit is included in
$A$ and its partner is included in $B$ yield a contribution
$\frac{2}{i (\epsilon_{A}^+ - \epsilon_{B}^-)}$ to $Z_{\rm off}$. 
All other possibilities to assign the
two orbits to $A, B, C, D$ can be treated in the same way.
Summation thus yields $2 F$, where

\begin{equation}
F \equiv    \frac{1}{i (\epsilon_{A}^+ - \epsilon_{B}^-)} -
 \frac{1}{i (\epsilon_{A}^+ - \epsilon_{D}^-)} -  \frac{1}{i (\epsilon_{C}^+ -
 \epsilon_{B}^-)} + \frac{1}{i (\epsilon_{C}^+ - \epsilon_{D}^-)}
\end{equation}

Upon taking derivatives and identifying the energies in a
columnwise way, Sieber/Richter pairs yield a non-oscillatory
contribution $\frac{1}{(i \epsilon)^3}$ to the complex
correlator, whereas the corresponding oscillatory term (obtained
after crosswise identification of energies) vanishes.

The simplest quadruplets beyond Fig. 1a and b are those with $m =
L - V  = 2$, i.e., either with two $2-$encounters or one
$3-$encounter. These quadruplets are shown in Fig. 2,
grouped into columns labeled by $1:1$, $1:2$, $1:3$, and $2:2$.
Here the first number counts the orbits {\it before} reconnection
(black lines) whereas the second number counts the orbits {\it after}
reconnection (colored or gray lines). 
To evaluate the resulting contributions, we have to sum over all ways of distributing these
orbits among $A$, $B$, $C$, and $D$ (placing the former orbits in $A$ or $C$
and the latter in $B$ or $D$ or vice versa) and apply the above diagrammatic rules. 

We have done so, 
and obtained the following result: All quadruplets of pseudo-orbits
that do not require time-reversal invariance (in the shaded rows) give
mutually canceling contributions to $Z_{\rm off}$,
and thus no off-diagonal contributions to the correlator. 

For time-reversal
invariant systems $Z_{\rm off}$ is non-zero.
When taking derivatives and identifying energies in a columnwise way, we see that
non-oscillatory terms in $C(\epsilon)$ arise only from quadruplets associated
to $1:1$, summing up to $\frac{3}{2(i\epsilon)^4}$.

In contrast, non-vanishing oscillatory contributions (obtained through the crosswise
representation) arise only from the quadruplets of type $2:2$. The first and
third of these diagrams in Fig. 2 mutually cancel. 
The leading off-diagonal
contribution to the oscillatory part can thus be attributed solely to
the second diagram, composed of two Sieber/Richter pairs
. 
For these quadruplets the sum over possible
assignments to $A$, $B$, $C$, and $D$ can be performed independently
for the two pairs.  We are therefore left with the square of the contribution of a
single Sieber/Richter pair, which must be divided by 2 since the
divisor $\prod_\lambda\omega_\lambda!$ in (\ref{zoff}) is now equal to 2.   
We thus obtain
a term $2F^2$ in $Z_{\rm off}$ and a contribution 
\begin{equation}
\frac{\partial^2(Z_{\rm diag} 2 F^2)}{\partial \epsilon_A^+
\partial \epsilon_B^-}{\Big|}_{\times} =
\frac{e^{2 i \epsilon}}{2 \epsilon^4}
\end{equation}
to $C(\epsilon)$,
coinciding with the leading oscillatory term in the 
RMT prediction (\ref{C}).

\begin{figure}[tbp]
\begin{center}
\includegraphics[scale=0.9]{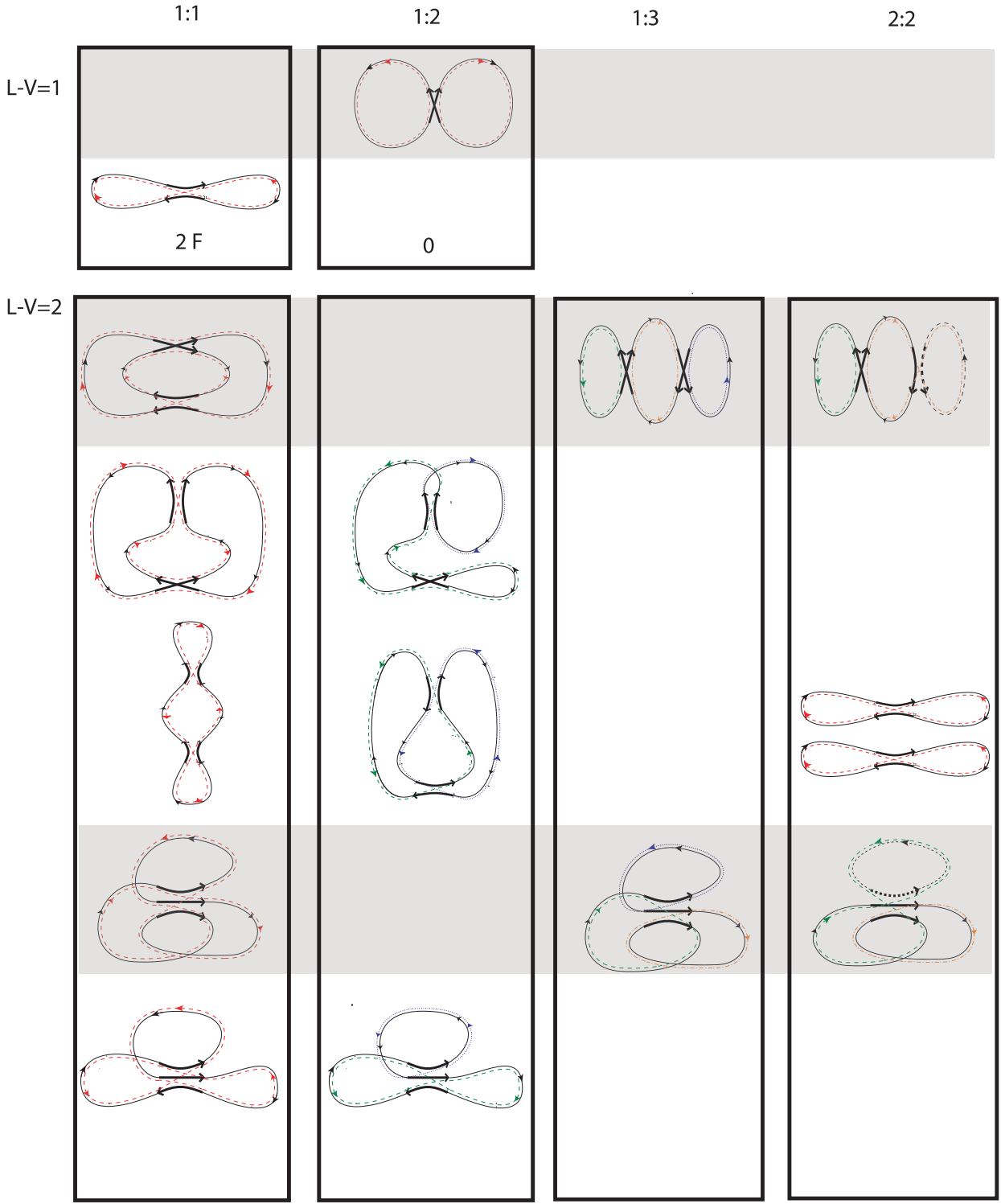}
\end{center}
\begin{minipage}[t]{17cm}
\caption{"Diagrams" responsible for the asymptotic expansion of
the complex correlator up to order ${\cal
O}(\frac{1}{\epsilon^4})$. The shaded rows contain pairs which
exist even in the absence of time-reversal invariance; the rest exists only
in the orthogonal case. The quadruplets in the first column are
marked by $1:1$ since they involve $one$ orbit before and after
reconnection. In  others columns, the number of orbits before and
after reconnection is $1:2, 1:3$ and $2:2$, respectively.
Pseudo-orbits containing a 3-encounter can have several partners
depending on the reconnection. The contributions to $Z_{\rm off}$
as defined in (\ref{zoff}) are shown explicitly for pairs with $L
= 2$ links and $V =1$ encounter. For $m = L - V = 2$, the
resulting contributions to $Z_{\rm off}$ are rather involved and
will be presented in a forthcoming publication.  }
\end{minipage}
\end{figure}


\begin{thebibliography}{99}
\bibitem{foot} In \cite{EssenFF,SR,Transport}, the term ``loop'' was
  used instead of ``link''; link is more appropriate for the majority
  of diagrams since most links start and end at different encounters.
\end{thebibliography}
\end{document}